\documentclass[conference]{IEEEtran}
\IEEEoverridecommandlockouts

\usepackage{url}
\usepackage{cite}
\usepackage{amsmath,amssymb,amsfonts}
\usepackage{algorithmic}
\usepackage{graphicx}
\usepackage{textcomp}
\usepackage{xcolor}
\usepackage{multirow}
\usepackage{listings}
\usepackage{balance}
\def\BibTeX{{\rm B\kern-.05em{\sc i\kern-.025em b}\kern-.08em
    T\kern-.1667em\lower.7ex\hbox{E}\kern-.125emX}}
\begin{document}

\title{Web-Based Simulator of Superscalar RISC-V Processors
\thanks{This work was supported by Brno University of Technology under project FIT-S-23-8141.}
}

\author{
   \IEEEauthorblockN{Jiri Jaros}
   \IEEEauthorblockA{\textit{Faculty of Information Technology} \\
                     \textit{Brno University of Technology}\\
                     Brno, Czech Republic\\
                     0000-0002-0087-8804}
\and
   \IEEEauthorblockN{Michal Majer}
   \IEEEauthorblockA{\textit{Faculty of Information Technology} \\
                     \textit{Brno University of Technology}\\
                     Brno, Czech Republic\\
                     misa@majer.cz}
\and
   \IEEEauthorblockN{Jakub Horky}
   \IEEEauthorblockA{\textit{Faculty of Information Technology} \\
                     \textit{Brno University of Technology}\\
                     Brno, Czech Republic\\
                     horkykuba@email.cz}
\and
   \IEEEauthorblockN{\IEEEauthorblockA{\hspace{7cm} Jan Vavra}}
   \IEEEauthorblockA{\hspace{7cm}\textit{Faculty of Information Technology} \\
                    \hspace{7cm} \textit{Brno University of Technology}\\
                    \hspace{7cm}Brno, Czech Republic\\
                    \hspace{7cm}jv369@seznam.cz}
}

\maketitle

\begin{abstract}
Mastering computational architectures is essential for developing fast and power-efficient programs. Our advanced simulator empowers both IT students and professionals to grasp the fundamentals of superscalar RISC-V processors, HW/SW co-design and HPC optimization techniques. With customizable processor and memory architecture, full C compiler support, and detailed runtime statistics, this tool offers a comprehensive learning experience. Enjoy the convenience of a modern, web-based GUI to enhance your understanding and skills.
\end{abstract}




\section{Introduction}
\label{sec:introduction}
In the rapidly evolving field of computer architecture, a deep understanding of superscalar processors is crucial for both IT students and professionals, particularly those focusing on writing high-performance and power-efficient code. However, mastering the intricacies of these architectures is challenging, especially when existing educational tools fall short. 

Current processor simulators are often either too complex and low level aiming at cycle accurate simulation of complex codes of yet non-existent processors, such as Intel Simcs Simulator \cite{intel_simics}, or lacking intuitive graphical interface, features such as supercalar out-of-order execution, processor customization, memory and cache hierarchy, or detailed runtime statistic. 

\subsection{State of the Art}
A comprehensive list of RISC-V simulators can be found on the RISC FIVE website \cite{riscv_simulators}. The Creator RISC-V RV32IMFD Online Assembly Simulator \cite{creator_simulator} is a powerful web-based tool that allows users to write, compile, and step through RISC-V RV32IMFD assembly code to observe program behavior. Its key features include processor and memory layout customization, runtime statistics collection, and online debugging. However, it only supports scalar processors and lacks a command-line interface (CLI) for benchmarking large program segments.

The Venus RISC-V Simulator \cite{venus_simulator} is a RISC-V instruction set simulator designed for educational purposes. It allows the simulation of more complex codes, but only on a scalar RISC-V processor, without the capability to inspect pipeline stages, hazards, or other detailed processor behaviors.

The Vulcan RISC-V Simulator for Education \cite{vulcan_simulator} offers several RISC-V instruction set extensions, along with side-by-side visualization of the program counter (PC), machine code, and original instructions, as well as register and memory visualization. However, it only supports a scalar core, and the web interface is still in the alpha stage.

Other notable simulators for RISC-V processors include Ripes \cite{RIPES} and Jupiter \cite{jupiter}. However, neither supports a superscalar pipeline or a web-based interface. 

In the search for inspiration in superscalar processor simulators, we must mention the excellent VSIM simulator \cite{vsim}, which our group has used for years in the Computer Architecture course. Developed in 2001, VSIM offers five architectures of superscalar processors from that era: Compaq Alpha 21264, Hewlett-Packard PA-8500, IBM Power3, Intel Pentium Pro/II/III, and MIPS R10000. VSIM allows partial customization of the processor architecture, the ability to load user-defined or random programs, and step-by-step simulation of program execution, including visualization of instruction and data flows between processor components. Unfortunately, this simulator is quite outdated and only runs on 32-bit Windows.

\subsection{Objectives}

The primary objective of the proposed web-based simulator is to bridge this educational gap by providing HPC developers with an accessible and illustrative tool to explore and understand the architecture of superscalar RISC-V processors. The simulator is designed to visually demonstrate each phase an instruction undergoes within the processor pipeline, allowing developers to identify potential bottlenecks and understand how different implementations of the same algorithm can impact runtime metrics such as execution time, cost or power consumption. By interacting with the simulator, developers can experiment with different processor configurations and observe their impact on runtime metrics.

Since the primary purpose of the simulator is educational, the initial version currently supports only the RV32IMFD instruction set. Future versions will add support for the 64-bit instruction set as well as vector extensions.

This hands-on approach aims to equip developers with the knowledge and skills needed to answer critical questions: Given an algorithm, how should one design a processor and optimize the code for the best performance, reasonable manufacturing cost and power consumption? By offering a~user-friendly interface and comprehensive support for customization and performance analysis, our simulator seeks to enhance the learning experience and prepare developers for the challenges of modern computing.


\section{Key Features of the Simulator}
\label{sec:features}

To address the challenges associated with understanding and teaching superscalar RISC-V processors, we have developed a comprehensive web-based platform-independent simulator.  Recognizing the need for both interactive and automated analysis, the simulator also includes a command-line interface (CLI) that allows for the analyzing of large programs in a batch processing manner, catering to advanced users who require more extensive testing capabilities. The proposed simulator offers following key features:

\begin{itemize}
  \item \textbf{User-Friendly Interface:} The simulator features an intuitive web interface that visually presents each block and instruction in the processor pipeline. It includes comprehensive documentation and tutorials, making it accessible for students and educators alike.

  \item \textbf{Fully Configurable Processors:} Users can customize various processor parameters, including fetch and issue width, size of register fields, reorder, load and store buffers, branch predictors implementations, number of functional units, supported operations and corresponding delays. The simulator also allows for detailed configuration of cache memory settings such as capacity, associativity, cache line size, and replacement strategy. This flexibility enables users to explore different processor designs and understand their impact on performance.
  
  \item \textbf{Forward and Backward Simulation:} The simulator supports both forward and backward instruction simulation, allowing users to step through the execution process in either direction. This feature aids in understanding the flow of instructions and the effects of architectural decisions on execution.

  \item \textbf{GCC Compiler Interface:} Integrated with the GCC compiler, the simulator enables users to compile C code into assembly, offering various optimization levels. The interface includes syntax highlighting and links between C and assembly code, helping users understand how different coding strategies impact low-level operations.

  \item \textbf{Comprehensive Runtime Statistics:} The simulator provides detailed performance metrics such as FLOPs, IPC, branch prediction accuracy, functional unit utilization, and cache hit rates. These metrics help users identify bottlenecks and optimize their code for better performance and efficiency.
  
  \item \textbf{Benchmark CLI}: For more advanced users, the simulator includes a command-line interface that allows for the benchmarking of complex programs in an automated, batch-processing manner.
  
  \item \textbf{Open Source}: The simulator's source code is available on GitHub, encouraging collaboration and allowing users to modify and extend the tool according to their needs. 
\end{itemize}

\subsection{Main Window with the Processor View}

The simulator's web interface is a multi-window application. Specific windows, including the main simulator window, code editor, memory editor, architecture settings, and runtime statistics, can be accessed from the left toolbar.



\begin{figure*}[p]
  \centering
  \rotatebox{90}{%
    \begin{minipage}{\textheight}
      \includegraphics[width=\textwidth, trim={0cm 0.0cm 0cm .0cm}, clip]{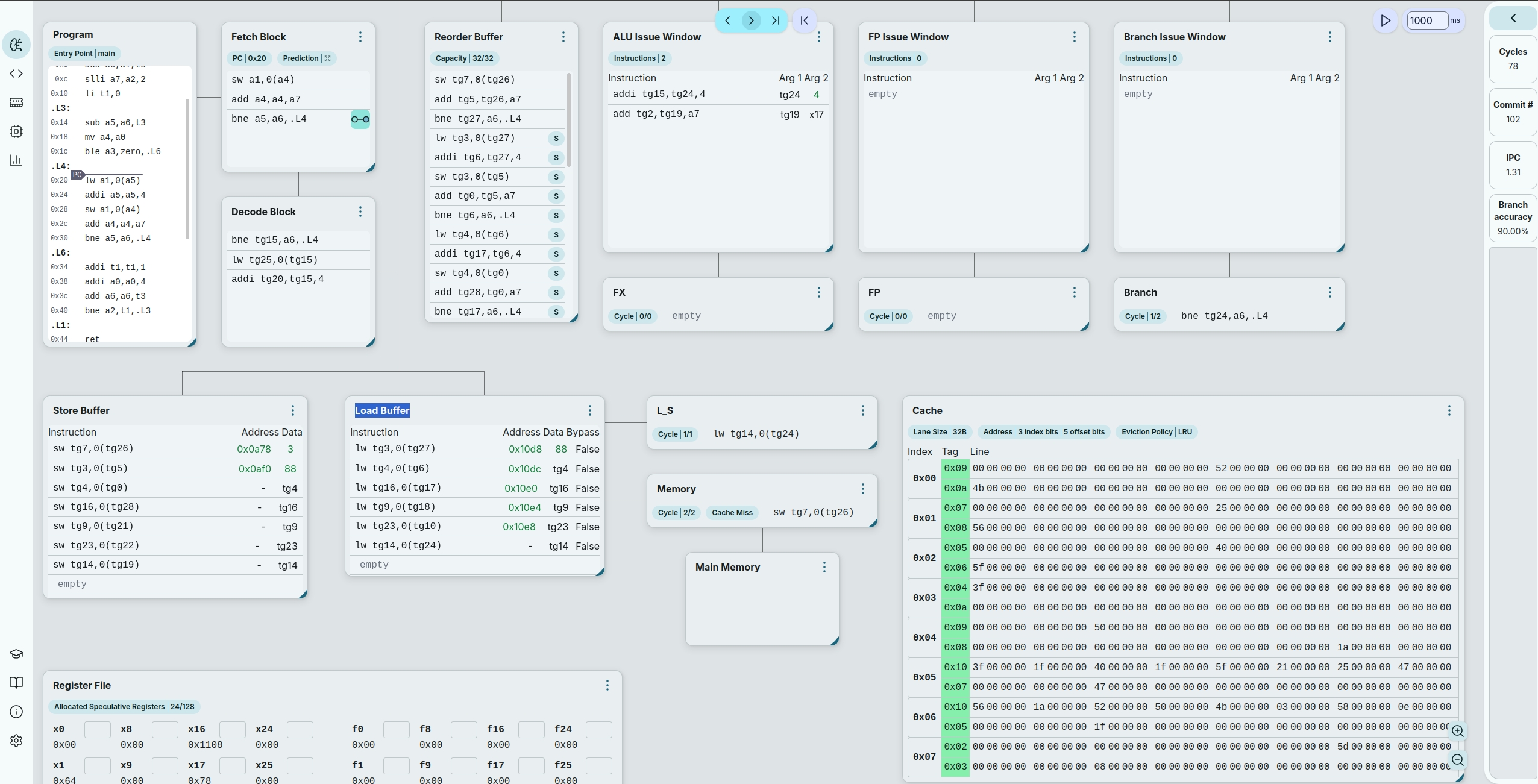}
      \caption{The main simulator window, displaying all processor components, including registers, cache, and main memory, with control buttons positioned at the top and basic statistics on the right.}
      \label{fig:main_window}
     \end{minipage}%
   }
\end{figure*}

The main simulator window, as shown in Fig. \ref{fig:main_window}, serves as the core interface of the simulator. It features processor schematics, a top simulation control bar, and a right-hand status bar. The processor schematics display essential components such as fetch and decode blocks, reorder (retire) buffer, and issue windows for the FX and FP ALUs, branch unit, and load/store (LS) components. Additionally, it includes a variable number of FX, FP, LS units, load/store buffers, and a memory unit connected to the cache. FX and FP registers are shown with their renamed tags and values, alongside the cache memory organized into lines. The simulation is fully controllable via mouse, keyboard, or on smartphones, with a slightly adjusted layout for mobile devices.

\begin{figure}[!b]
  \centering
  \includegraphics[width=\columnwidth, trim={0cm 0.0cm 0cm .0cm}, clip]{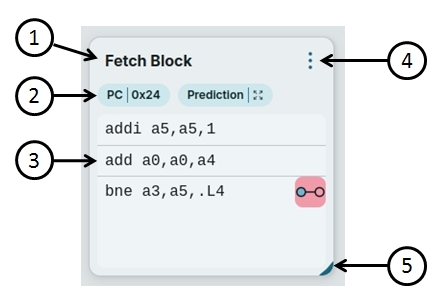}
  \caption{Graphical representation of the fetch block with (1) block name, (2) simulation information, (3) active instructions, (4) pop-up details, and (5) resize bar." }
  \label{fig:control_elements}
\end{figure}

All blocks share the same control elements, as shown in Fig. \ref{fig:control_elements}. The top left corner displays the name of the block (1), while the top right corner features a button (4) that opens a pop-up window with detailed information about the current status of the block. The second line (2) provides the most crucial real-time information about the block. The bottom right corner (5) allows for resizing the block. The remaining area of the block (3) is specific to its function, typically containing a list of active instructions and their details, such as the state of the branch predictor, actual names of the registers, valid bits, etc.

\begin{figure}[!bt]
  \centering
  \includegraphics[width=\columnwidth, trim={0cm 0.0cm 0cm .0cm}, clip]{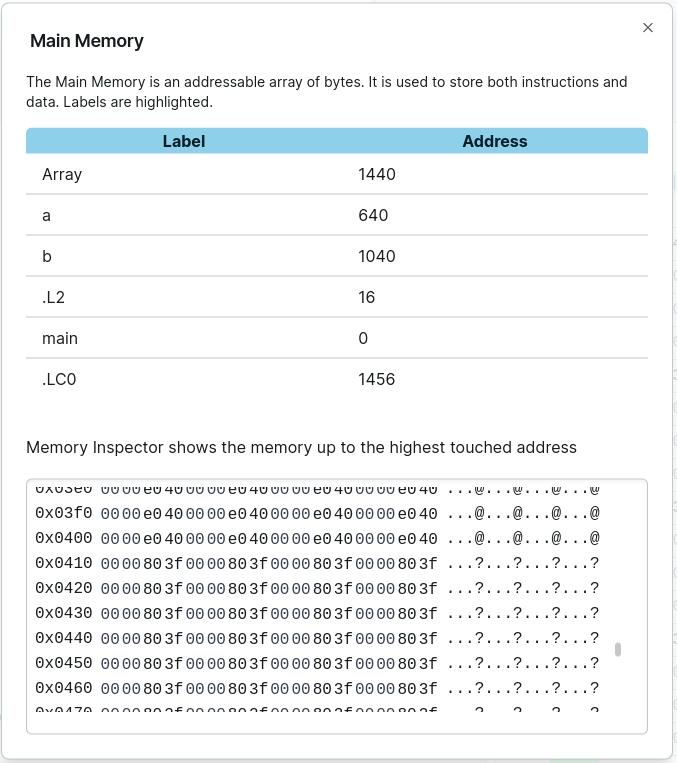}
  \caption{A pop-up window displaying the current state of the memory, including allocated arrays, their starting addresses, and a memory dump." }
  \label{fig:mem_details}
\end{figure}

The schematic view of the simulation state offers a general overview but lacks detailed information. For additional details, pop-up previews are available. Clicking on a block or instruction opens a window displaying relevant data in a~tabular format. For instructions, this includes timestamps of key phases (fetch, decode, ...), parameter values, and flags (e.g., validity flag). The details are  block specific; for instance, the main memory block (as shown in Fig. \ref{fig:mem_details}) reveals all program pointers, their addresses, and an expanded view of the entire memory.

\begin{figure}[!tb]
  \centering
  \includegraphics[width=\columnwidth, trim={0cm 0.0cm 0cm .0cm}, clip]{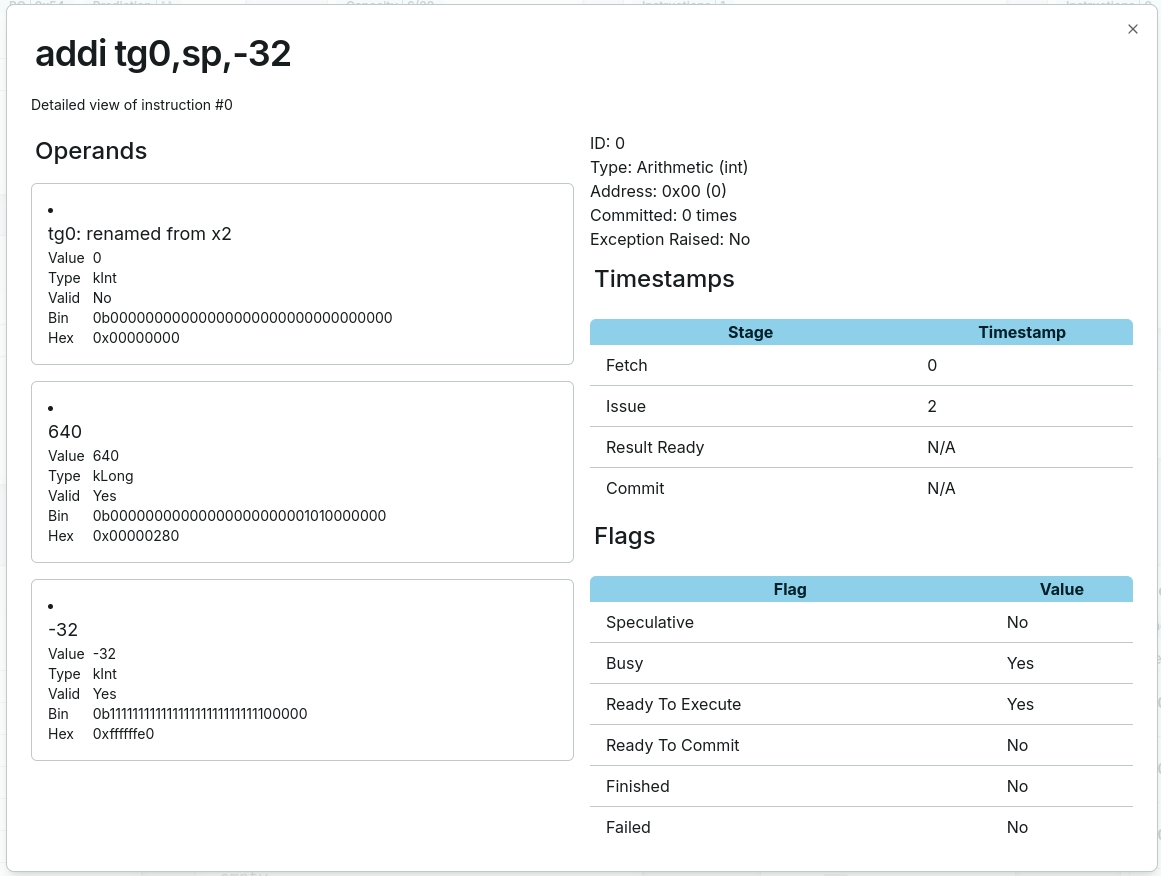}
  \caption{A pop-up window that displays instruction current state, parameters, renaming details, values and validity, along with instruction flags and the timestamps of phase completions.}
  \label{fig:instr_details}
\end{figure}

Hovering over an instruction or register highlights all its occurrences across other blocks, making it easier to comprehend the simulation state. Additionally, hovering over an instruction parameter reveals a tooltip with its value, and for registers, information about their renaming is also displayed. Finally, clicking on an instruction opens a pop-up window with detailed information about its state, as shown in Fig. \ref{fig:instr_details}.

The right-hand panel displays selected statistics and the debug log. It has two states; default and expanded. In the default view, it shows the number of cycles, committed instructions, IPC, and branch prediction accuracy, while the expanded view includes additional metrics such as FLOPs and cache hit rate. Complete statistics are available on a separate page (see Sec. \ref{sec:perf_statistics}). Each log message is timestamped with the cycle in which it was generated, and clicking on the message number navigates the simulation to that specific cycle. 

\subsection{Code Editor}
\label{sec:code_editor}

\begin{figure}[!b]
  \centering
  \includegraphics[width=\columnwidth, trim={0cm 0.0cm 0cm .0cm}, clip]{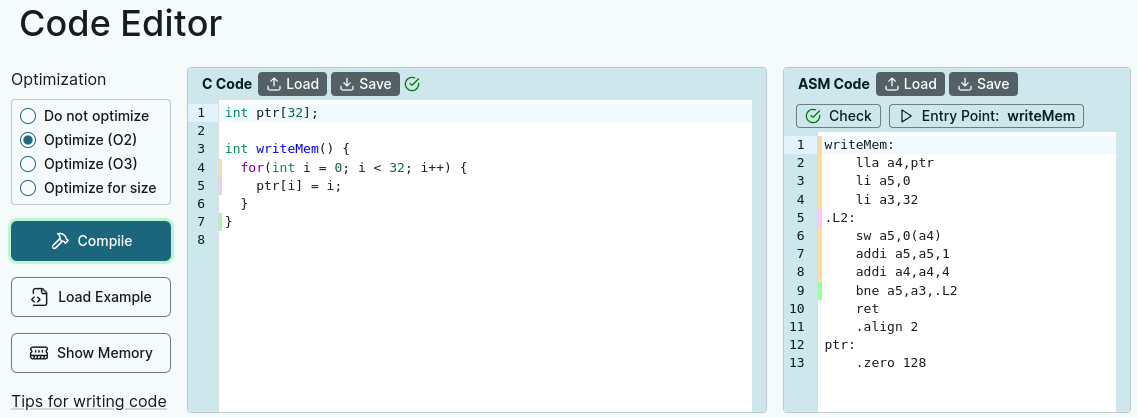}
  \caption{Code editor displaying C and Assembly codes, with compiler parameters and control buttons.}
  \label{fig:code_editor}
\end{figure}

\begin{figure}[!tb]
  \centering
  \includegraphics[width=\columnwidth, trim={0cm 0.0cm 0cm .0cm}, clip]{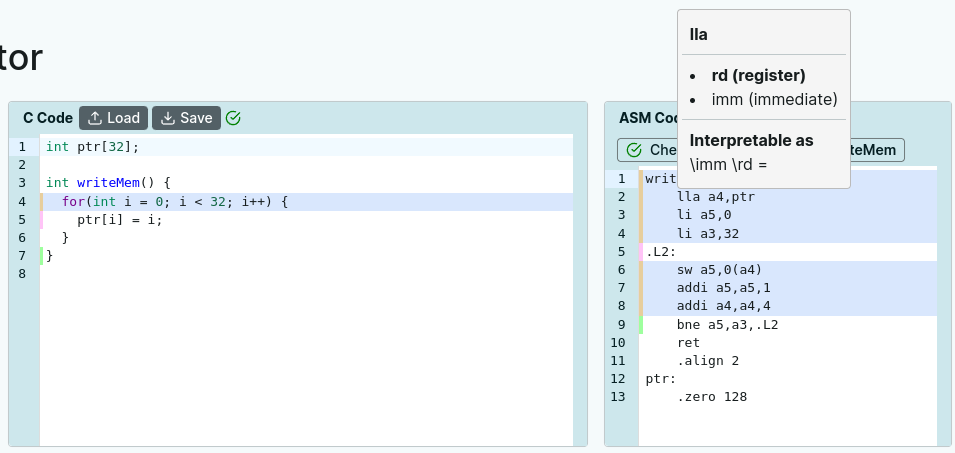}
  \caption{Code editor showing the link between C and Assembly codes, with instruction details displayed in a bubble window.}
  \label{fig:code_hover}
\end{figure}

The code editor allows users to input programs in both C and RISC-V assembly languages, see Fig. \ref{fig:code_editor}. The entry point can be set to the first instruction or any specified label. When the code is entered in C, it can be compiled into assembly using four optimization levels. In this case, the C and assembly codes are linked through highlighting, enabling visualization of how C statements are translated into assembly instructions, see Fig. \ref{fig:code_hover}. If the code requires global arrays, the C language keyword \texttt{extern} can be used, and the array contents can be filled in the Memory Settings window (see Sec. \ref{sec:memory_and_processor_config}). In assembly code, users can use labels such as \texttt{.word} to define memory arrays. Users can also load basic assembly and C examples or load and save complex code from and to files.

\begin{figure}[!tb]
  \centering
  \includegraphics[width=\columnwidth, trim={0cm 0.0cm 0cm .0cm}, clip]{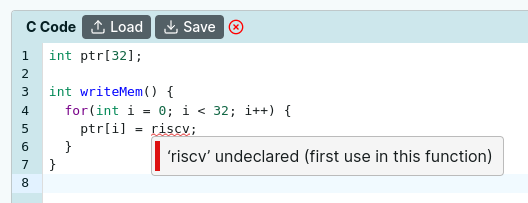}
  \caption{Syntax error visualization in the C code.}
  \label{fig:c_error}
\end{figure}

\begin{figure}[!tb]
  \centering
  \includegraphics[width=0.8\columnwidth, trim={0cm 0.0cm 0cm .0cm}, clip]{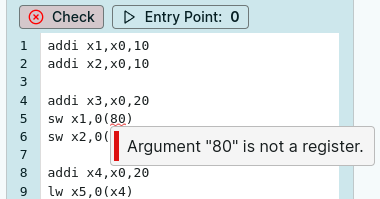}
  \caption{Syntax error visualization in the Assembly code.}
  \label{fig:asm_error}
\end{figure}

The editor is implemented using the CodeMirror library\footnote{\url{https://codemirror.net/5/}}, which provides a robust user interface with features such as syntax highlighting, keyboard shortcuts, line numbering, error highlighting (Fig. \ref{fig:c_error} and \ref{fig:asm_error}), and more.

\subsection{Memory and Processor Architecture Settings}
\label{sec:memory_and_processor_config}

\begin{figure}[!tb]
  \centering
  \includegraphics[width=\columnwidth, trim={0cm 0.0cm 0cm .0cm}, clip]{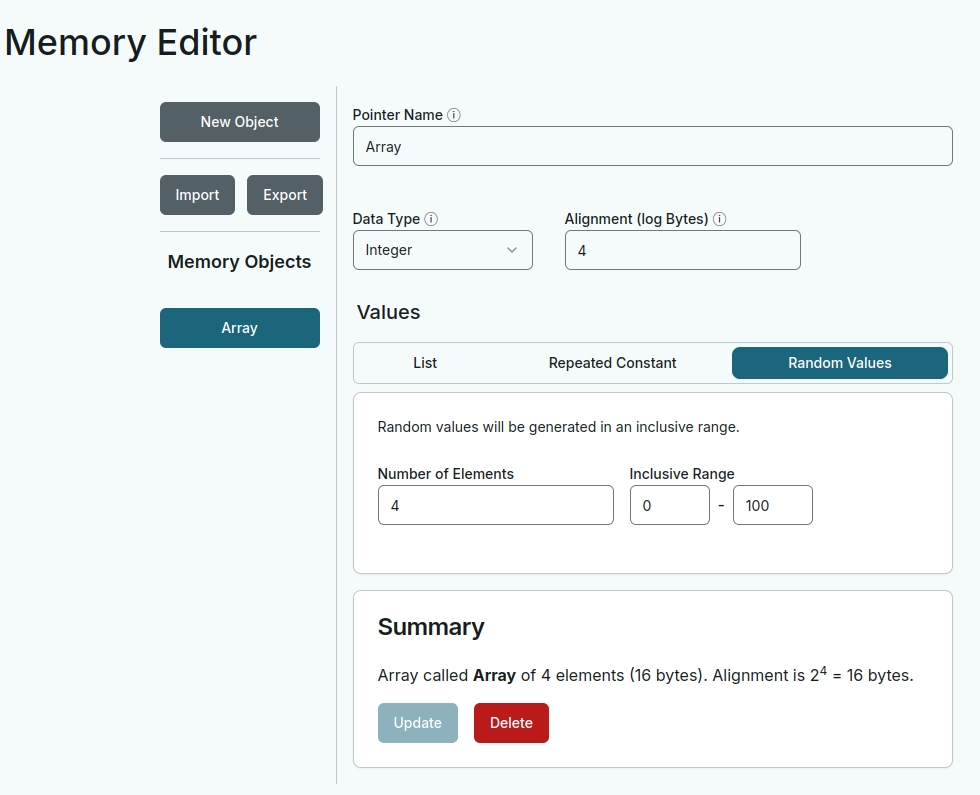}
  \caption{Memory editor allowing to define static global arrays and fill them with user data.}
  \label{fig:memory_editor}
\end{figure}

The Memory Settings window allows users to populate memory with custom data, see Fig. \ref{fig:memory_editor}. Users can define static global arrays of various basic data types and specify their alignment. Arrays can be populated with user-specified values separated by commas, repeated constants (e.g., zeros), or random values. Additionally, memory dumps can be imported and exported in binary or CSV format.

\begin{figure}[!tb]
  \centering
  \includegraphics[width=\columnwidth, trim={0cm 0.0cm 0cm .0cm}, clip]{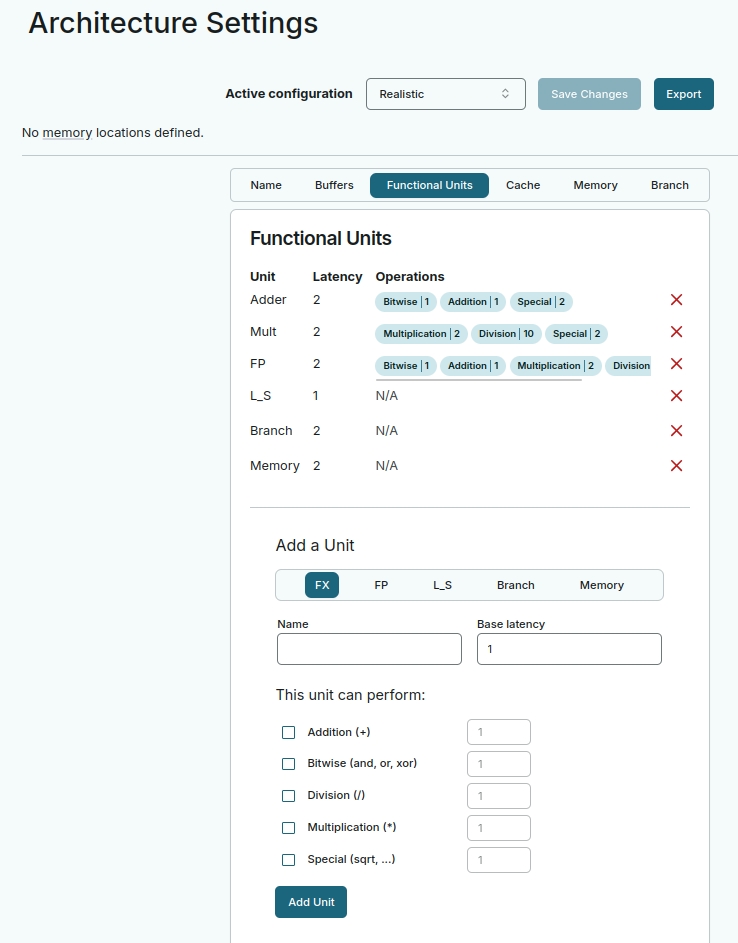}
  \caption{Architecture settings for custom processor and memory configuration, including processor width, functional units, cache organization, memory subsystem, and branch predictor.}
  \label{fig:architecture_settings}
\end{figure}

The Architecture Settings window enables users to customize the processor architecture and cache in detail, see Fig. \ref{fig:architecture_settings}. The window is organized into several tabs, each grouping related settings. At the top, users can switch between different architectures and import or export configurations using JSON files.

The first tab allows users to name the architecture and set the core and memory clock speeds in Hz. The second tab, titled Buffers, controls the superscalar processor's width by adjusting the reorder buffer size, the number of instructions fetched and committed per cycle, flush penalty, and the number of jumps the fetch unit can handle within a single cycle. The third tab addresses the functional units, categorized into FX, FP, LS, branch, and memory. FX and FP units can vary in supported instructions and associated latencies, while LS, memory and branch units allow for latency specification only.

The Cache tab provides options to enable or disable the L1 cache, define the number of cache lines, their size, and associativity. Users can also choose a replacement policy from LRU, FIFO, or Random, and determine the store behavior, either write-back or write-through. Additionally, users can control the cache line replacement delay and cache access delay. The Memory tab allows for the configuration of the load and store buffer size, load and store latency, call stack size, and the register rename file size. Finally, the Branch prediction tab lets users set the branch target buffer size, pattern history table size, predictor type (zero, one, or two-bit), predictor default state, and choose between local or global history shift registers.

\subsection{Runtime Statistics}
\label{sec:perf_statistics}

\begin{figure}[!bt]
  \centering
  \includegraphics[width=\columnwidth, trim={0cm 0.0cm 0cm .0cm}, clip]{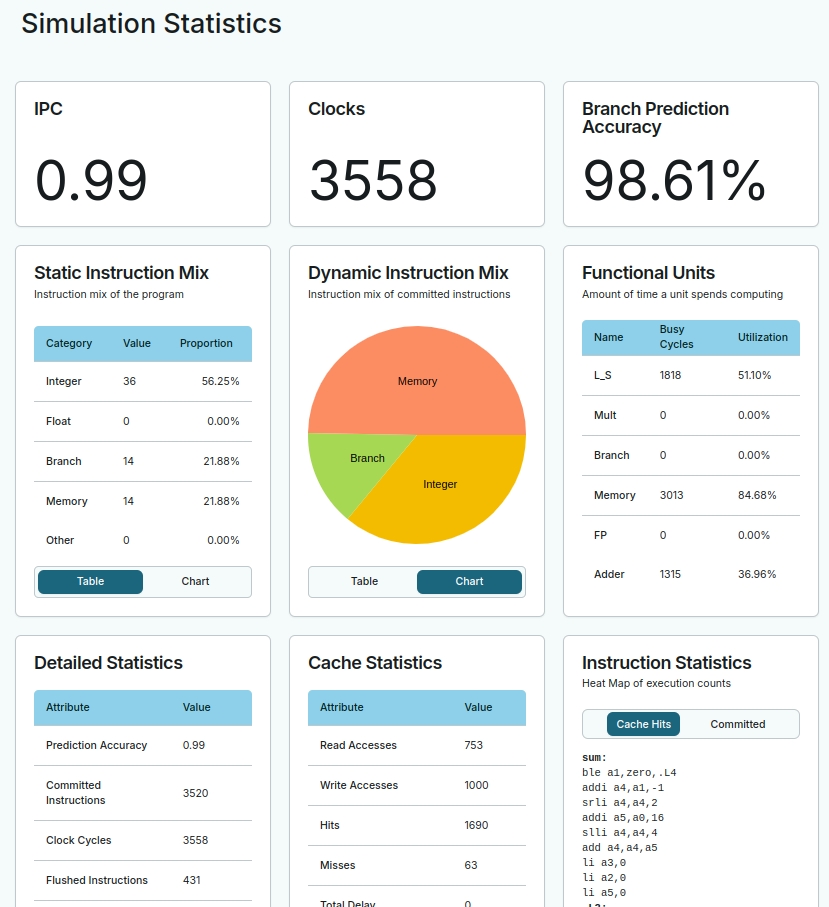}
  \caption{Collected runtime statistics cover a wide range of performance statistics. }
  \label{fig:statistics}
\end{figure}

The Runtime Statistics window provides detailed useful insight into the code execution, including both static and dynamic instruction mix in tabular and graphical formats. It also summarizes the number and percentage of busy cycles for each unit, as well as cache statistics, including the number of accesses, hit and miss ratios, and bytes written. Additionally, the window displays various detailed metrics such as predictor accuracy, total executed cycles, total number of committed instructions, number of reorder buffer flushes, FLOPS, IPC, wall time, and many more, see Fig. \ref{fig:statistics}.

\subsection{Command-Line Interface}

The Command-Line Interface (CLI) allows users to execute large programs written in C or assembly language and collect runtime statistics. The CLI requires two mandatory arguments: the assembly language source code in a text file and the architecture description in JSON format. Additional parameters allow to specify the program's entry point, memory configuration, data dump, and various levels of output verbosity and format (either text or JSON). The CLI must be connected to the server using host and port parameters, with an optional connection to the GCC compiler.

\section{Implementation and Deployment}
\label{sec:implementation}

The proposed web-based simulator is a client-server application with two client interfaces: a command-line interface (CLI) and a web-based interface. Both clients interact with the server API to present simulation results, while all simulation logic is handled server-side. The web client is developed in JavaScript, using the React library\footnote{\url{https://react.dev/}} and the Next.js framework\footnote{\url{https://nextjs.org/}}. It communicates with the simulation server via HTTP, using a JSON-based API. The HTTP server is implemented with the Undertow library\footnote{\url{https://undertow.io/}}. HTTPS support is provided through an NGINX proxy server\footnote{\url{https://nginx.org}}. The global state, facilitating communication between modules, is managed using sessions wrapped in React context and maintained by the Redux library\footnote{\url{https://redux.js.org/}}.

\subsection{Simulator Architecture and Simulation Step Manager}

The server-side, which encompasses the entire simulator logic, is written in Java and utilizes the JavaFX library. The simulator is organized into modules, taking full advantage of object-oriented design principles. The central class \texttt{BlockScheduleTask} maintains the list of references on all simulated blocks in the processor, using the observer design pattern to broadcast the changes in the processor state. 

The simulator's memory is represented as a 1D byte array with a predefined capacity. Memory modules operate in a~transactional mode. Functional blocks that request data from memory generate an object representing a transaction. Upon registration, memory management populates this object with information about the transaction's completion time. Transactions enable easy configuration of memory access times, support cache line flushing, and include metadata useful for interactive simulation.

The initialization of the simulation involves several steps, including configuration validation, loading of register and instruction definitions, initialization of statistics and the simulation step manager, parsing of the assembly program, memory setup, construction of all processor components, initialization of the register file with specific values, and setting the PC register to the program's entry point.

The simulation can proceed either clock cycle by clock cycle (step by step) or run continuously to completion. Each simulation clock cycle (a step) is executed by sequentially calling all blocks, which are arranged in a queue based on their position in the pipeline. The simulation of functional units is divided into two sub-steps to allow the completion of the current instruction and the loading of the next one within a single clock cycle. It is important to note that the functional blocks currently do not support internal pipelining, which is commonly used in components such as the floating-point ALU. The simulation ends when the pipeline is empty or when the stack pointer reaches the bottom of the call stack, indicating process completion as the main routine is exited. 

The simulation step manager also collects runtime statistics.

\subsection{Register Representation and Instruction Interpretation}

Registers are represented as 64-bit arrays, even though the simulator currently supports only 32-bit instructions. The interpretation of their values depends on the type of instruction being executed. Each register also contains metadata defining the data type in use, making code debugging in the GUI more intuitive by displaying the intended value (e.g., a char) instead of a raw bit array. Additionally, registers maintain all necessary information for renaming. Each register tracks the number of references; architectural registers use a list of all renamed copies, while renamed (speculative) registers hold a pointer to the corresponding architectural register. This setup allows for detailed tracking of register renaming.

The simulator fully supports the RV32I instruction set with the M and F extensions, including pseudo-instructions and directives (e.g., \texttt{.word}). However, privileged instructions and instructions for context switching are not supported, as the simulator does not run an operating system. Branching and memory instructions have been modified to work with indices into arrays representing code and data memory segments, rather than with memory addresses.

The instruction set is defined in a configuration JSON file and can be easily extended, see Listings \ref{lst:instruction_definition}.

\begin{lstlisting}[caption={Definition of the add instruction, its parameters and interpretation.},label={lst:instruction_definition},numberstyle=\tiny\color{gray},numbers=left,
    numbersep=5pt,showspaces=false,captionpos=b, basicstyle=\ttfamily\small] 
{
  "name": "add",
  "instructionType": "kArithmetic",
  "arguments": [
    {
       "name": "rd",
       "type": "kInt",
       "writeBack": true
    },
    {
       "name": "rs1",
       "type": "kInt"
    },
    {
        "name": "rs2",
        "type": "kInt"
    }
  ],
  "interpretableAs": "\rs1 \rs2 + \rd ="
},
\end{lstlisting}

The execution of an instruction is managed by the \texttt{Expression} class, which implements a simple stack-based interpreter using postfix notation, as shown in Listing \ref{lst:instruction_definition} under \texttt{interpretableAs}. The interpreter also handles operands directly encoded in the instruction opcode, such as the PC register in jump and branch instructions. The output of an expression may be twofold: the first possible output is the value that remains on the stack after the interpretation is executed, a mechanism used by expressions to calculate jump addresses or conditions. The second possible output is the assignment to a variable within the expression. The binary operator \texttt{=} in the expression has a side effect, writing the value into the register. Exceptions are generated during code execution (e.g., when accessing an unauthorized address, division by zero). The existence of an exception is checked when the instruction is committed.

The simulator also supports backward simulation, enabling users to inspect changes in the processor state in detail. This is implemented as a forward simulation with $t-1$ clock cycles. While this approach significantly simplifies the implementation, it requires the simulation to be deterministic and imposes higher computational demands on the server. Backward simulation is only available in the web application, where it is intended for use with small programs running over a few thousand clock cycles.

\subsection{Compiler Integration}

The simulator utilizes the GCC compiler's cross-compilation to translate C programs into RISC-V assembly. When the code is ready for compilation, the web client packages the source code and sends it to the server via a POST request. The server then generates a shell script to execute the compiler, collecting the compiled assembly program along with a log of any potential compiler errors. The result is sent back to the web client. 

For processing the compiled or user defined assembly code, a two-pass approach was chosen. In the first pass, instructions and memory definitions (directives) are processed. The program text is divided into language units (tokens such as symbols, comments, or new lines). The tokens are then processed sequentially in a loop according to the grammar, and the individual instructions are stored. Instruction objects are linked by references to objects that describe their behavior and to register objects.

Listing \ref{lst:memory_labels} shows examples of memory definitions in the assembly program. The compiler supports the following directives: \texttt{.byte}, \texttt{.hword}, \texttt{.word}, \texttt{.align}, \texttt{.ascii}, \texttt{.asciiz}, \texttt{.string}, \texttt{.skip} a \texttt{.zero}.

\begin{lstlisting}[caption={Example of memory definitions in the assembly language. Memory defined in this way is referenced in the program using labels, e.g. \texttt{arr}.}, captionpos=b,label={lst:memory_labels}, numberstyle=\tiny\color{gray},numbers=left, numbersep=5pt,showspaces=false,captionpos=b, basicstyle=\ttfamily\small]
   x:
     .word 5 # integer variable x

     .align 4
   arr:
     .zero 64 # 64 bytes with 16B alignment

   hello:
     .asciiz "Hello World" # null-terminated 
                           # string
\end{lstlisting}

After the first pass, not all operand values are defined, because an operand may refer to a label that has not yet been processed. The second pass fills in the missing values, completing the program processing. A complication, when filling in the values, is the support for arithmetic expressions in instruction arguments (e.g., \texttt{lla x4, arr+64}). This feature is implemented because the compiler often generates such expressions. Therefore, memory allocation takes place between the first and second pass. After allocation, all label values are known, and the final values of instruction arguments can be calculated. Jump instructions use relative values for jumps, so it is sometimes necessary to subtract the instruction's position from the absolute value of the label. Expressions are evaluated by a simple evaluation program, which must have access to the label values.

Finally, the assembler output may contain a large amount of information that is redundant for the simulator and also reduces the readability of the code. Therefore, the compiler output is passed through a filter that removes unnecessary directives, labels, and data.

Once the program is compiled, it is loaded into memory. User data defined on the memory settings page (and with the \texttt{extern} keyword in the code) must be statically allocated in memory. The allocation is performed with respect to the data type and alignment requirements. The compiled code then operates with labels set to the starting addresses of specific memory arrays. Since the compiler's application binary interface (ABI) requires a call stack, the memory initialization process allocates space for the stack at the beginning of the memory and stores the corresponding pointer in register x2 (the stack pointer). User data is then stored after the stack.

\subsection{Deployment}

Continuous integration and development is managed under the GitHub version control system. Deployment is carried out using three Docker containers. The first container, called \texttt{simserver}, is responsible for all the simulation logic, while the second container, called \texttt{web}, mediates requests between the client application and the simulation server using web technologies. Both containers open a single external communication port. The third container, \texttt{nginx\_proxy}, provides HTTPS encapsulation for internal HTTP communication using the supplied server SSL certificate. The containers can be compiled and deployed automatically using the \texttt{docker-compose} command.


\section{Testing and Evaluation}
\label{sec:evaluation}

The simulator is quite extensive piece of software with almost 33 thousand lines of code. During the implementation of the simulator, unit testing was intensively used and gradually expanded.  The project in its current state contains more than 400 unit tests. The code coverage in the simulator is 83\%, and the coverage in the simulator's blocks is 94\%.

The system was tested as a whole from many aspects. Each instruction has its own test to verify its correct behavior. This type of test typically checks the state at the end of the simulation. Additionally, the test script ensures that all provided code samples run on the simulator without errors. The functionality of several more complex programs is also tested, such as array sorting using the quicksort algorithm, working with a linked list, and polymorphism (dynamic dispatch).

The web application was manually tested using Google Chrome and Mozilla Firefox. The web interface was also evaluated with Google Lighthouse\footnote{\url{https://developer.chrome.com/docs/lighthouse/overview}}. Performance tests showed that rendering typically takes around 80 ms.

The simulator also underwent user testing. Twenty participants, consisting of IT students and faculty members, were asked to complete several tasks and evaluate the user experience. The simulator achieved an average score of 8.4/10. The testing also revealed several minor bugs, which were subsequently fixed.

\subsection{Performance Evaluation}

For performance evaluation, the simulator was profiled in server mode. Additionally, a simple benchmark was developed using the Java Microbenchmark Harness (JMH)\footnote{\url{https://www.baeldung.com/java-microbenchmark-harness}}.

The most important conclusion from the performance testing is the following: in server mode, about 60\% of the request handling time is consumed by working with the JSON format. This format is inherently unfavorable for performance. As a result, the dominance of the communication overhead means that further performance gains from optimizing the simulation itself are diminishing. Exploring a change in the communication protocol is a direction worth investigating in future work.

Load testing was also conducted using Apache JMeter\footnote{\url{https://jmeter.apache.org/}}. The characteristics of the test are as follows:
\begin{itemize}
  \item two test sizes: 30 and 100 users,
  \item each user interactively simulates 40 steps of the simulation for one of two programs,  
  \item ramp-up time of 4 seconds, with a 1-second pause between each user's request (think time),
  \item use of gzip.
\end{itemize}

Using gzip compression increased throughput on the local server by 40\%. Table \ref{tab:perf_data} presents the measured data. All measurements were conducted locally on a laptop with an Intel i5 8300H processor and 16 GB of DDR4 RAM. The conclusion is that the server handles a smaller number of concurrent users well, regardless of the runtime mode, although Docker has a noticeable impact on application performance. A larger number of users significantly affects latency, degrading the user experience. During the test, there were no application crashes or query failures. In real-world operation, latency is likely to be higher due to the longer travel distance of packets over the internet. A larger number of users can be managed by running the application on more powerful hardware or by distributing the load across multiple servers.

\begin{table}[]
\caption{The measured latency values for the four specified scenarios.}
\centering
\begin{tabular}{|l|r|rr|r|}
\hline
\multirow{2}{*}{Mode}  & \multicolumn{1}{l|}{\multirow{2}{*}{\#users}} & \multicolumn{2}{l|}{Latency [ms]}                                & \multicolumn{1}{l|}{\multirow{2}{*}{Throughput [trans/s]}} \\ \cline{3-4}
                        & \multicolumn{1}{l|}{}                                 & \multicolumn{1}{l|}{Median} & \multicolumn{1}{l|}{90th percentile} & \multicolumn{1}{l|}{}                                   \\ \hline
\multirow{2}{*}{Direct} & 30                                                    & \multicolumn{1}{r|}{70,66}  & 118                                & 25.96                                                   \\ \cline{2-5} 
                        & 100                                                   & \multicolumn{1}{r|}{680}    & 1248.9                             & 53.61                                                   \\ \hline
\multirow{2}{*}{Docker} & 30                                                    & \multicolumn{1}{r|}{77}     & 283                                & 24.49                                                   \\ \cline{2-5} 
                        & 100                                                   & \multicolumn{1}{r|}{1135}   & 2031.9                             & 42.07                                                   \\ \hline
\end{tabular}
\label{tab:perf_data}
\end{table}

\balance

\section{Conclusions}
\label{sec:conclusions}

The proposed web-based simulator for superscalar RISC-V processors is a substantial contribution to computer architecture education and research. By offering an accessible and interactive platform, it facilitates a deeper understanding of complex processor architectures and encourages experimentation and innovation.

To the best of our knowledge, this is the most advanced web-based simulator for a superscalar RISC-V processor with L1 cache support, designed for educational use, benchmarking code snippets, and architectural evaluation. The simulator boasts several significant features. It offers a user-friendly graphical interface, with fully configurable processor components, allowing users to export and share configurations. The simulator supports C and assembly language programs, various levels of code optimization through the integrated GCC compiler, and an intuitive code editor. Performance analysis is enabled by a range of built-in metrics, and for more complex programs, the command-line interface allows batch evaluation of different algorithm versions, facilitating direct comparisons across different architectures.

The intended users of this simulator are primarily IT students specializing in processor design and HPC computing. We believe that a deep understanding of processor architecture will contribute to the development of highly optimized RISC-V processors for custom applications. Additionally, understanding how processors handle specific code patterns and snippets will lead to better optimization of HPC codes across various architectures. We hope that even advanced HPC developers will benefit from this simulator by evaluating different implementations and processor configurations.

The simulator will be used in the upcoming academic year during the Computation Systems Architectures course at the Faculty of Information Technology, Brno University of Technology. Nearly 250 students will leverage its features to solve assignments focused on optimizing specific code patterns concerning the provided architecture.

Future work will focus on expanding the simulator's capabilities. Several directions are under consideration. One direction is to enhance the processor architecture with additional features, such as vector units, advanced branch predictors, pipelined functional units, or deeper cache hierarchies. Another potential area of development is improving the code development and simulation environment by adding breakpoints, watches, dynamic memory allocation, atomic operations, and more. Additionally, runtime statistics could be expanded to measure the chip area consumed by specific blocks based on their complexity or estimate the processor's power consumption using realistic manufacturing technology.

\begin{figure}[!ht]
  \centering
  \includegraphics[width=0.4\columnwidth, trim={0cm 0.0cm 0cm .0cm}, clip]{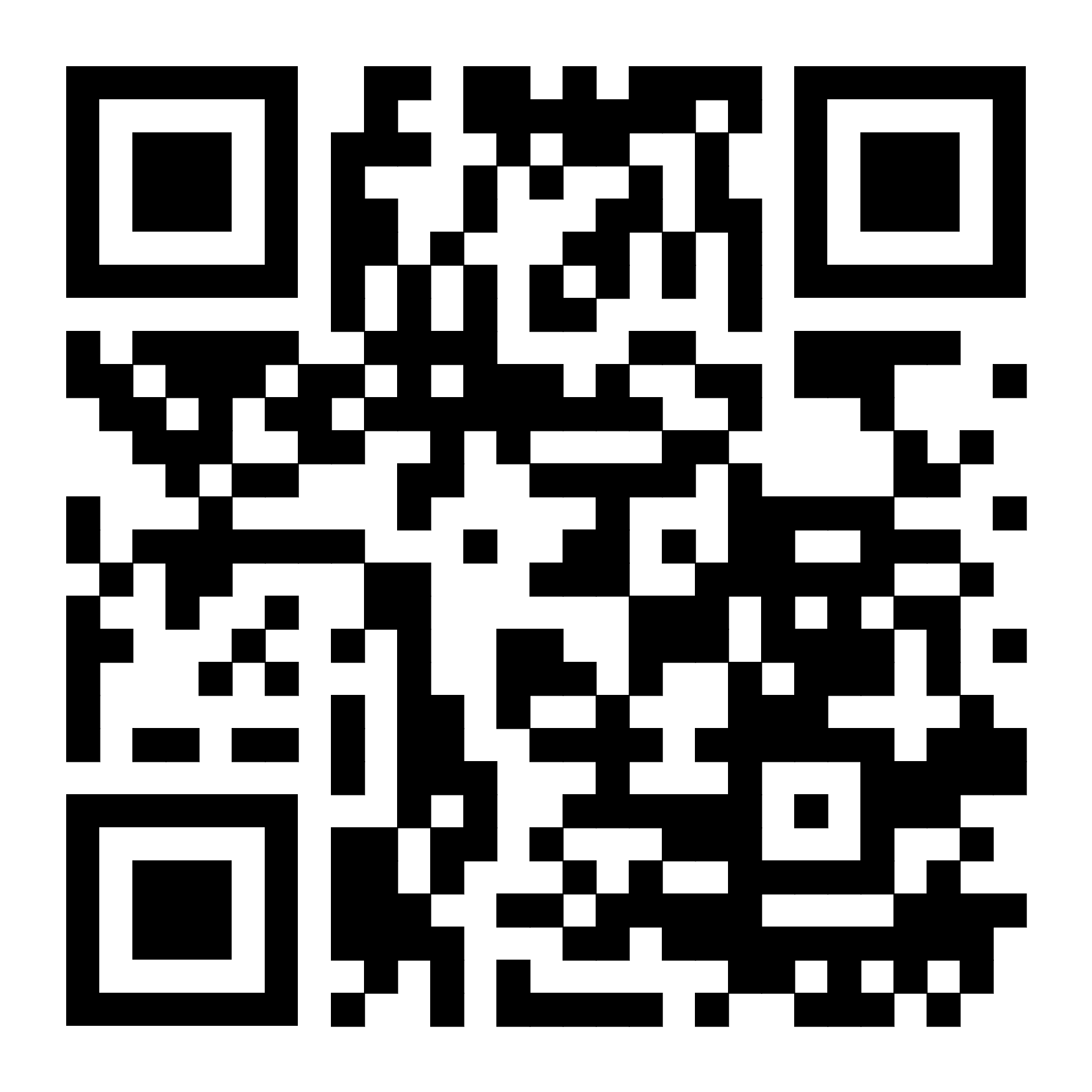}
  \includegraphics[width=0.4\columnwidth, trim={0cm 0.0cm 0cm .0cm}, clip]{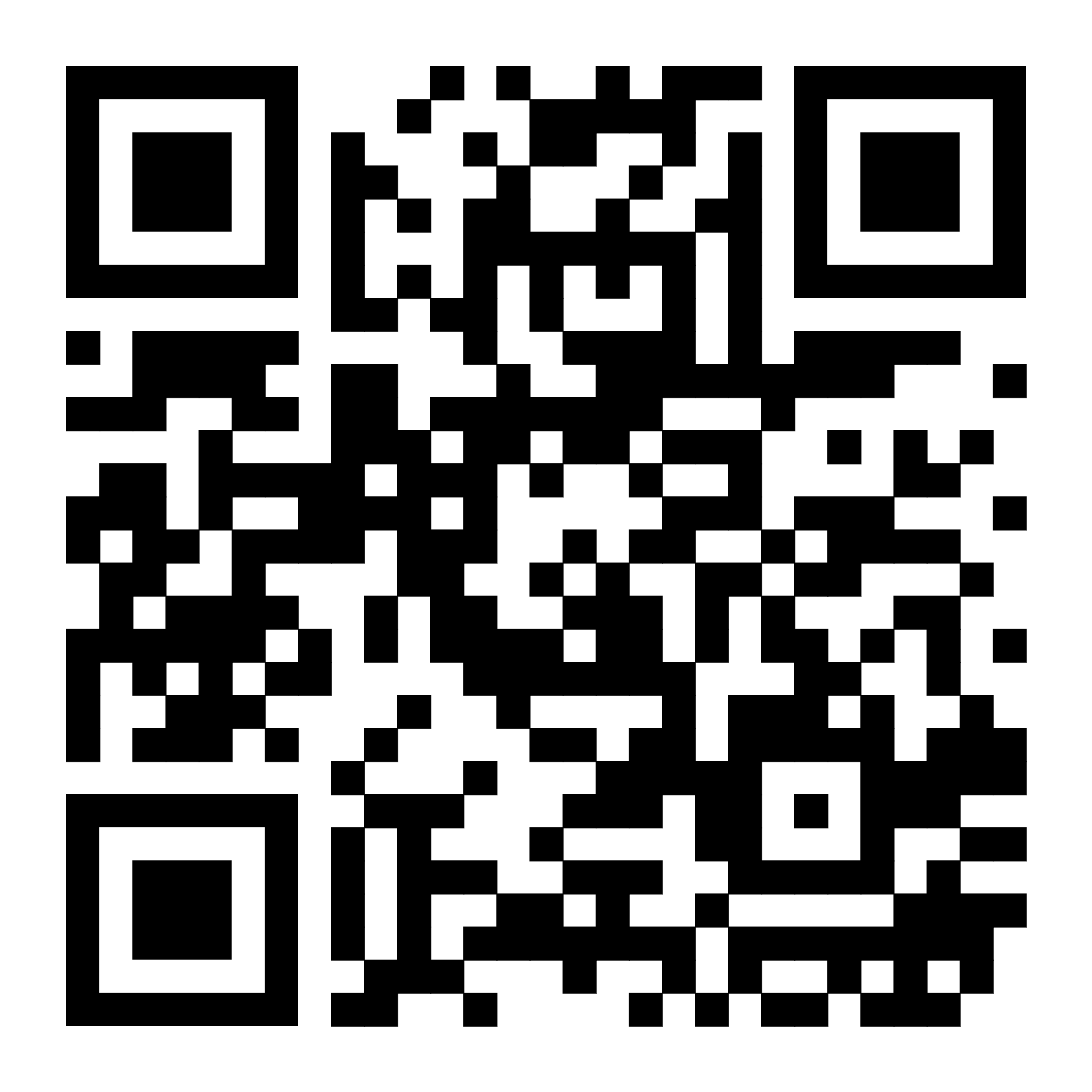}
  \caption{QR codes with source codes on GitHub and live demo.}
  \label{fig:QR_codes}
\end{figure}

Finally, we provide links to the source code on GitHub and a live instance of the simulator, as shown in Fig. \ref{fig:QR_codes}.

\bibliographystyle{IEEEtran}
\bibliography{IEEEabrv,Sections/Bibliography}

\end{document}